\documentstyle[12pt]{article}
\input epsf

\def\like{{\cal L}}
\def\wml{w^{\cal L}_\theta}
\def\eql#1{\label{eq:#1}}
\def\ec#1{\ref{eq:#1}}
\def\vx#1{\vec x_{#1}}
\def\be{\begin{equation}}
\def\ee{\end{equation}}
\def\bea{\begin{eqnarray}}
\def\eea{\end{eqnarray}}

\def\fun#1#2{\lower3.6pt\vbox{\baselineskip0pt\lineskip.9pt
        \ialign{$\mathsurround=0pt#1\hfill##\hfil$\crcr#2\crcr\sim\crcr}}}

\def\ksection{\arabic{section}}

\def\thesection{}

\begin{document}
\thispagestyle{empty}
\renewcommand{\thefootnote}{\fnsymbol{footnote}}

\font\ssqfont=cmssq8 scaled 2500
\font\sqfont=cmssq8 scaled 1100
\null\vspace*{-104pt}
\begin{center}
\parbox{8.0in}{
\epsfxsize=48pt \epsfbox{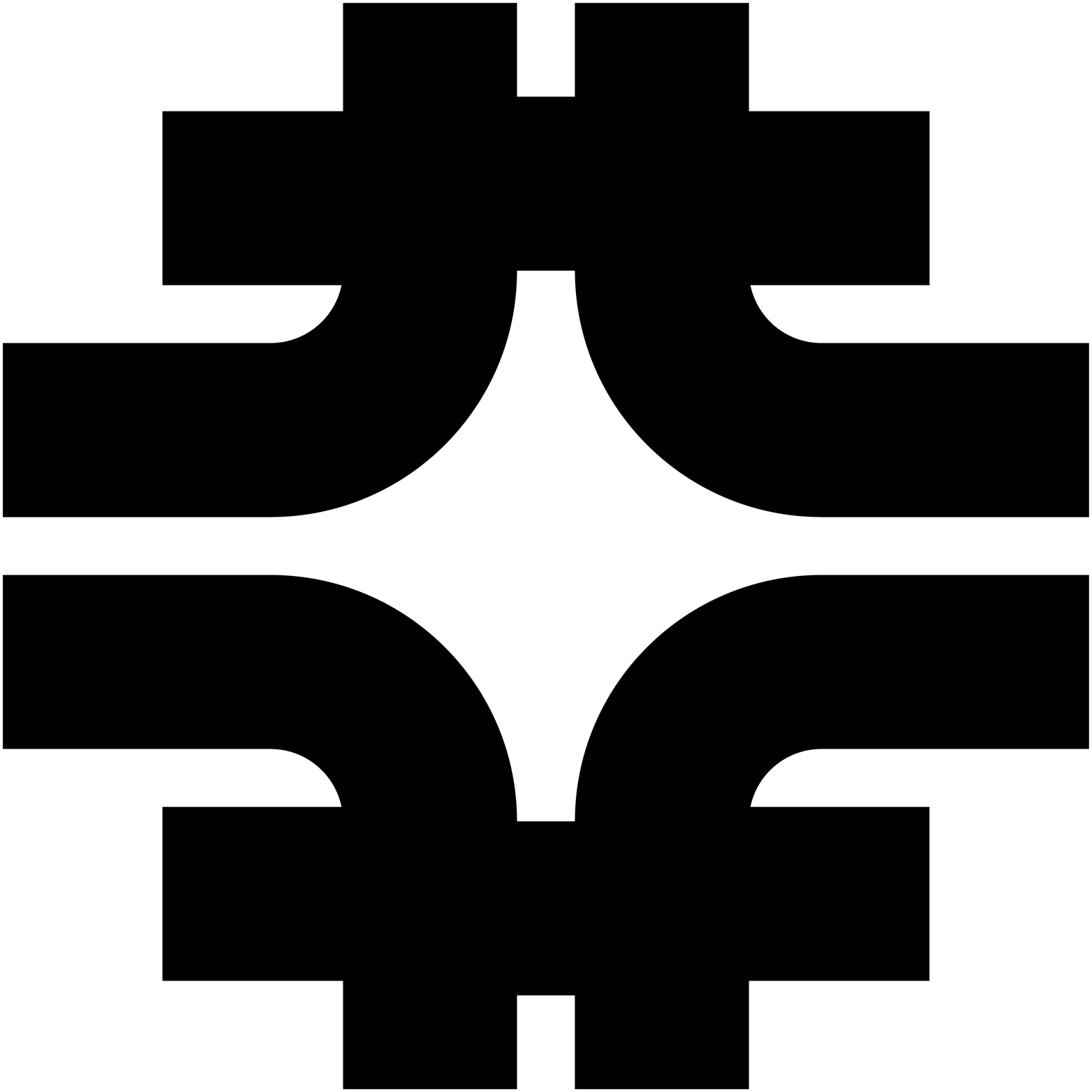}
\vspace*{-33pt}\hspace*{72pt}{\ssqfont Fermi National Accelerator Laboratory} 
}
\end{center}
\begin{flushright}
{\footnotesize
\today}
\end{flushright}
\nopagebreak
\vspace*{8.5in}\centerline{
\epsfxsize=18pt \mbox{\epsfbox{Logo.eps} \ \  $^{\mbox
{\sqfont{Operated by Universities Research Association Inc.\ 
under contract with the United States Department of Energy}}}$} }
\vspace*{-8.75in}
\baselineskip=24pt

\begin{center}
{\Large \bf Likelihood Analysis of Galaxy Surveys}\\
\vspace{1.0cm}
\baselineskip=14pt

Scott Dodelson\footnote{Electronic mail: {\tt dodelson@hermes.fnal.gov}}\\
{\em NASA/Fermilab Astrophysics Center \\
Fermi National Accelerator Laboratory, Batavia, IL~~60510}\\
\vspace{0.4cm}
Lam Hui\footnote{Electronic mail: {\tt lhui@fnal.gov}}\\
{\em NASA/Fermilab Astrophysics Center \\
Fermi National Accelerator Laboratory, Batavia, IL~~60510}\\
\vspace{0.4cm}
Andrew Jaffe\footnote{Electronic mail: {\tt jaffe@physics.berkeley.edu}}\\
{\em  Center for Particle Astrophysics, 301 LeConte Hall\\
University of California,\\
Berkeley, CA~~94720}
\end{center}

\baselineskip=18pt

\begin{quote}
  \hspace*{2em} One of the major goals of cosmological observations is
  to test theories of structure formation. The most straightforward way
  to carry out such tests is to compute the likelihood function $\like$,
  the probability of getting the data given the theory. We write down
  this function for a general galaxy survey.  The full likelihood
  function is very complex, depending on all of the $n$-point functions
  of the theory under consideration. Even in the simplest case, where
  only the two point function is non-vanishing (Gaussian perturbations),
  $\like$ cannot be calculated exactly, primarily because of the Poisson
  nature of the galaxy distribution. Here we expand $\like$ about the
  (trivial) zero correlation limit. As a first application, we take the
  binned values of the two point function as free parameters and show
  that $\like$ peaks at $(DD - DR + RR)/DD$. Using Monte Carlo techniques, we
  compare this estimator with the traditional $DD/DR$ and Landy \&
  Szalay estimators. More generally, the success of this expansion
  should pave the way for further applications of the likelihood
  function.

\vspace*{12pt}


\end{quote}


\newpage
\baselineskip=18pt
\setcounter{page}{2}
\addtocounter{footnote}{-3}

\thesection{\centerline{\large \bf I. INTRODUCTION}}
\setcounter{section}{1}
\setcounter{equation}{0}
\vspace{18pt}

Recently, there has been a vast increase in both the quality and quantity
of data accumulated by observational cosmologists. The high
quality of this observational work sets a standard for theorists 
to interpret the data as carefully and meaningfully as possible.  For
these purposes, the likelihood function has become a powerful tool for
all cosmologists.  The likelihood function, $\like$, is the probability of
a given data set given a theory; as such it is a natural way of
connecting observations with theory. Examples of its application to
cosmological data sets include cosmic microwave background anisotropies,
gravitational lens studies and fitting absorption lines in QSO spectra.

Until now, though, there has been no attempt to apply likelihood
techniques to galaxy surveys. Instead, one typically determines the
two-point function, $\xi(r)$, or the power spectrum, $P(k)$, from the
survey using appropriately chosen estimators (usually depending on the
distribution of observed pairs of points).  On the one hand, this is
quite surprising: it would be wonderful to use all the information in a
survey instead of just one small subset of it.  If the galaxy
distribution was Gaussian, so that only the two-point function mattered,
it would make sense to neglect all the other information in the
catalogue. The galaxy distribution, however, is decidedly non-Gaussian,
so the likelihood function, which incorporates information from all the
$n$-point functions, seems more appropriate. On the other hand, the
likelihood function for galaxies in a survey is a very complicated
beast. Even if the underlying density field were Gaussian (so that only
the two point function was non-zero), we would not necessarily expect an
ad hoc estimator to be ideal. Further, and perhaps most importantly,
the galaxies themselves are a Poisson (or other) sample of this
underlying field, which introduces non-Gaussianity into the distribution
of galaxy locations. In \S II, we write down $\like$ and show that 
even for the Gaussian case it is computationally unfeasible to calculate.

In \S III, we propose one way to extract some information
from the likelihood function: we expand $\like$ about
its value when all correlations are zero. This weak
correlation limit proves very fruitful. As an example, we
consider a general 2-D (angular) survey (although the validity of our
results is not confined to 2-D) and a theory
with the free parameters being the value of the two-point
correlation function $w_\theta$ in different angular bins.
Thus, in this example, $\like$ is a function of the data and
the parameters $w_\theta$. We show that, for each $\theta$,
$\like$ peaks at
\be
\wml = {DD(\theta) - DR(\theta) + RR(\theta) \over DD(\theta) }
\eql{EST}
\ee
where $DD,DR,$ and $RR$ are respectively the number of
pairs of particles in the data set with angular distances
within the bin denoted by $\theta$; the number of pairs
of particles---one in the data set and one in a random
catalogue---in the $\theta$ bin; and the number of random
pairs in the bin $\theta$. This estimator is very close
to the one introduced by Landy \& Szalay (1993). (For other
estimators, see Peebles (1980), Hewett (1982), and Hamilton (1993); 
for a recent discussion of the Landy \& Szalay estimator and
a generalization to higher order correlations, see 
Szapudi \& Szalay (1997)).
We regard this as a success of the expansion: we are able to
extract a reasonable estimator, one shown to be significantly
more effective than the traditional $DD/DR$. In \S IV, we
analyze these estimators more carefully using Monte Carlo simulations to
see which has the lowest variance. These simulations suggest that
the maximum likelihood estimator of Eq.~\ec{EST} has a smaller variance
than both the traditional estimators and the Landy \& Szalay
estimator. In the course of performing these simulations, we
have uncovered additional terms in the variance of the Landy-Szalay
estimator which dominates over the normal Poisson term when there
are many galaxies in the survey. We present a simple
derivation of these additional terms in Appendix B.

While the application developed in \S III and \S IV is useful, we
feel the most important feature of our analysis is the realization
that the likelihood function can be approximated in a 
meaningful way. This opens the door to a host of applications, some
of which we speculate about in the conclusion.
Finally, to improve the readability of the text, we have
shifted most of the calculational details of the
expansion in \S III to Appendix A.


\vspace{24pt}
\thesection{\centerline{\large \bf II. The Likelihood Function}}
\setcounter{section}{2}
\vspace{18pt}
The probability of finding $N$ galaxies at positions
${\vec x_1, \vec x_2, \ldots, \vec x_N}$ and no galaxies
elsewhere in the survey volume $V$ is given by 
\begin{eqnarray}\label{white}
\like &\equiv& P[\vec x_1, \vec x_2, \ldots, \vec x_N; \Phi_0(V) |
w_2(\theta), n] 
= \exp\{W_0\} \ 
n^N dV_1 dV_2 \cdots dV_N \cr
&\times
&
\Bigg[ W_1(\vec x_1) W_1(\vec x_2) \cdots W_1(\vec x_N)  \cr
&+ & W_2(\vx1,\vx2) W_1(\vx3) \cdots W_1(\vx{N}) + {\rm permutations} \cr
&+ & \ldots
\cr
&+ & W_2(\vx1,\vx2) \cdots W_2(\vx{N-1},\vx{N}) + {\rm permutations} \Bigg]
\end{eqnarray}
where $\Phi_0(V)$ is the ``proposition'' that there are no galaxies in
the sample other than those at the specified ${\vec x_i}$.
Here we have assumed that there are no higher order correlations. 
The general expression including higher order correlations was first
written down by White (1979); we do not reproduce it here.
So the free parameters in the ``theory'' are the density $n$ and
the two point correlation function, $w_2(\theta)$. The $W_i$'s
are then given simply by
\begin{eqnarray}
W_0 &=& -n V + {n^2\over 2} \int dV_1 dV_2 w_2(\vx1,\vx2)
\cr
W_1(\vx1)  &=& 1 - n \int dV_2 w_2(\vx1,\vx2)
\cr
W_2(\vx1,\vx2) &=& w_2(\vx1,\vx2)
\end{eqnarray}
Here the infinitesmal ``volume'' $dV_i$ surrounds the position
$\vx{i}$ and can be either 2-D (areas) or 3-D (volumes). For a realistic
survey, all of these volume integrals are to be weighted by the
selection function of the survey: this replaces the probability that
there are galaxies at the ${\vec x_i}$ and no others with the
probability of {\em observing} galaxies at the ${\vec x_i}$ and nowhere
else. In principle, by suitably recomputing the probability of the
statement $\Phi_0(V)$, complications such as redshift-space distortions
could be included.

We should also note that this expression assumes a particular model for
the relationship between galaxy positions and the underlying ``number
density field'': the galaxies are a {\em Poisson sample} of the density.
That is, the probability of finding a single galaxy in an infinitesimal
volume $\delta V$ around ${\vec x}$ is just $n({\vec x})\delta
V\ll1$. We can then use a suitable biasing prescription to connect the
number density field $n({\vec x})$ to the mass density field.

Equation \ref{white} is deceptively simple. A given line
contains $m$ occurences of the two point function $W_2$.
The deceptiveness lies in the phrase ``$+$ permutations'':
there are many, many terms included in this phrase. Consider
the line with $N/8$ $W_2$'s. The number of ways of choosing
$3N/4$ galaxies and assigning each a factor of $W_1(\vx{i})$
is 
\be
{N! \over (3N/4)! ~ (N/4)!
}.
\nonumber
\ee
The remaining $N/4$ galaxies can 
be arranged into $W_2$'s in 
\be
{ (N/4)! \over 2^{N/8}~ (N/8)! }
\nonumber 
\ee
ways. So the total number of terms on this one
line is
\be
{ N!~ 2^{-N/8} \over (N/8)!~ (3N/4)!}   \simeq
( 8N )^{N/8}
\ee
where the approximate equality uses 
Stirling's formula. Even for a small catalogue with a
thousand galaxies, this line contains $10^{488}$
terms! Each of these must be calculated separately;
and there are $N/2$ lines with comparable numbers of
terms (and this is just for the Gaussian case!). Clearly,
an exact calculation of the likelihood function is
out of the question. 
We need to develop approximation schemes that will enable
us to circumvent an exact calculation.

\vspace{48pt}
\thesection{\centerline{\large \bf III. Weak Correlation Limit}}
\setcounter{section}{3}
\vspace{18pt}

We are especially interested in the regime where the correlations are
weak ($w\ll1)$. In this regime, we might further expect non-Gaussianity
to be negligible, at least when we start from an initially Gaussian
density field as from inflationary theories.

\vspace{24pt}
\thesection{\centerline{\bf A. Zero Order Solution}}
\setcounter{section}{3}
\vspace{18pt}

We want to expand Eq.~\ref{white}\ about $w_2=0$. So let us first find
the likelihood when $w_2=0$.
In this case the likelihood function is extremely simple; it reduces
to the Poisson distribution:
\begin{equation}
P[\vec x_1, \vec x_2, \ldots, \vec x_N; \Phi_0(V)] 
= \exp\{-n V\} \ 
n^N dV_1 dV_2 \cdots dV_N 
\end{equation}
If we differentiate this with respect to the
one free parameter, the density $n$, we find that
the likelihood function peaks
when 
\begin{equation}
n = \bar n = N/V.
\end{equation}
The width of the likelihood function gives a measure of
how accurately the parameter $n$ is known. One way to
estimate this width is to expand $\ln(\like)$
around $n=\bar n$:
\be
\ln(P) = \ln(P[\bar n]) + {1\over 2} {d^2 \ln P \over
dn^2}|_{n=\bar n} (n - \bar n)^2 + \ldots
\ee
In this simple example,
\begin{eqnarray}
\ln(P) &&= -n V + N\ln(n) + {\rm constant}
\cr
&& = -\bar n V + N\ln(\bar n) + {1\over 2}\left(
{-N \over \bar n^2}\right) (n - \bar n)^2 + \ldots
\end{eqnarray}
So we can identify the width of the likelihood
function as $(\bar n^2/N)^{1/2} = (N/V^2)^{1/2}$, so that
$\delta n/n = N^{-1/2}$, which is of course the correct answer
for a Poisson distribution. Note that the Poisson distribution
considered as a function of $n$ is skewed. So if we choose as our
estimator (say) the mean rather than the mode, we would find a different
answer; for a large survey, this is clearly a tiny effect.

We now pursue this approach including
correlations. Specifically, we calculate the first
derivative of $\ln(P)$ and set it to zero in order
to determine the free parameters. Then we calculate the
second derivative to calculate the width of the
likelihood function.
 
\vspace{24pt}
\thesection{\centerline{\bf B. First Order Solution for $n$}}
\setcounter{section}{3}
\vspace{18pt}

The starting point is Eq.~\ref{white}:
\begin{eqnarray}\label{lnone}
\ln P
= W_0 &+& N\ln n 
+ \ln\Bigg[ \prod_a W_1(\vx{a})   + {1\over 2} \sum_a \sum_b^a 
W_2(\vx{a},\vx{b}) \prod_c^{a,b} W_1(\vx{c})  \cr
&+& {1\over 2^3} \sum_a \sum_b^a \sum_c^{a,b} \sum_d^{a,b,c}
W_2(\vx{a},\vx{b}) W_2(\vx{c},\vx{d}) \prod_e^{a,b,c,d} W_1(\vx{e})
\Bigg]
\end{eqnarray}
where we have dropped irrelevant constants [such as $\ln V$].  We need
to expand this to second order in $w_2$; then when differentiating to
find the maxmium, we will get a linear equation for $w_2$.  Therefore,
only terms up to second order in $W_2$ have been kept. Another facet
of this expansion---which is detailed in Appendix A---is that all the
$W_1$'s in the $W_2 W_2 \prod W_1$ term (last in the square brackets
above) can be set to one. Similarly, all but one of the $W_1$'s in terms
of the type $W_2 \prod W_1$ can be set to one, etc.  We have introduced
the notation of superscripts on $\sum,\prod$. These indicate which
galaxies should {\it not} be summed or mulitplied over. Thus, $\sum_a^b$
means sum over all $a=1,\ldots,N$ except $a=b$. For example,
\begin{equation}
  \sum_a^{b,c}\equiv \sum_a(1-\delta_{ab})(1-\delta_{ac}),
\end{equation}
the generalization to more or fewer superscripts should be plain.

Now, a word about the factors of $2$, which may be a source of
confusion: the term with one $W_2$ clearly has one factor of two to
account for the fact that we are double counting $W_2(\vx1,\vx2)$ and
$W_2(\vx2,\vx1)$.  The term with two $W_2$'s obviously needs two of
these types of factors. But it also needs another one to account for
$W_2(\vx1,\vx2)W_2(\vx3,\vx4)$ and $W_2(\vx3,\vx4)W_2(\vx1,\vx2)$, hence
the $1/2^3$ factor before the last term.

It is worthwhile here to introduce the notation of Landy
\& Szalay. For these purposes we divide the survey volume
into $K$ cells, each of which is so small that it contains
at most one galaxy. Then
\be
\int dV = {V\over K} \sum_{i=1}^K
\ee
where the sum over $i$ includes even those cells that don't have
galaxies in them.  With this notation, $W_0$ and $W_1$ can be rewritten
as
\be
W_0 = -n V + {n^2 V^2\over 2K^2} \sum_{i,j} w_2(\vx{i},\vx{j})
\end{equation}
\begin{equation}
W_1(\vx1) = 1 - {n V\over K} \sum_i w_2(\vx1,\vx{i})
\end{equation}
It is worth noting that if one goes over the original derivation by
White (1979), the summation over $i$ and $j$ in the above two
equations should, strictly speaking, range over only empty cells.  
We do not place such restrictions here, essentially working in the
continuum limit, even though we represent the integrals as discrete
sums. 

Now let's find the value of the density at the
maximum of the likelihood function. We
need to differentiate Eq.~\ref{lnone}\ with respect to $n$.
\begin{eqnarray}\label{lnnone}
{\partial\ln P \over \partial n}
&=& {\partial W_0\over \partial n} + {N\over n} \cr
&+& \Bigg[ \prod_a W_1(\vx{a})   
+ {1\over 2} \sum_a \sum_b^a 
W_2(\vx{a},\vx{b}) \prod_c^{a,b} W_1(\vx{c})  \cr
&+& {1\over 2^3} \sum_a \sum_b^a \sum_c^{a,b} \sum_d^{a,b,c}
W_2(\vx{a},\vx{b}) W_2(\vx{c},\vx{d}) 
\Bigg]^{-1}
\cr
&\times&
{\partial \over \partial n}
\Bigg[ \prod_a W_1(\vx{a})   
+ {1\over 2} \sum_a \sum_b^a 
W_2(\vx{a},\vx{b}) \prod_c^{a,b} W_1(\vx{c})  \Bigg]
\end{eqnarray}
Here we have used the fact that $\partial W_2/\partial n = 0$.
To go further we need the other partial derivatives:
\be
{\partial W_0 \over \partial n} = 
- V + {nV^2\over K^2}\sum_{i,j} w_2(\vx{i},\vx{j})
\qquad ;\qquad
{\partial W_1(\vx{a}) \over \partial n} = 
- {V\over K}\sum_i w_2(\vx{i},\vx{a})
\ee
Note first that setting
Eq.~\ref{lnnone}\ to zero will give a solution of the form
$n = N/V + O(w_2)$. We therefore need to keep only terms linear
in $w_2$. 
After differentiating the numerator, it will be of
order $w_2$. We keep only these linear terms and set the denominator to
one.
Thus, we are left with
\begin{eqnarray}\label{lnntwo}
{\partial\ln P \over \partial n}
&=& - V + {N\over n} + {nV^2\over K^2}\sum_{i,j} w_2(\vx{i},\vx{j})
- 
 {V\over K}\sum_{a,i} w_2(\vx{i},\vx{a}) \\ \nonumber
&=& - V + {N\over n} + {V\over K}\sum_i\left[
  {nV\over K}\sum_j w_2(\vx{i},\vx{j})-\sum_a w_2(\vx{i},\vx{a})\right]
\end{eqnarray}
The latter two terms differ only through a scaling and whether the sums
range over the observed galaxies or the random catalogue; in the absence
of clustering they will be equal and cancel. Hence, they will differ
only by another factor of $w_2$; to this order in $w_2$, they therefore vanish.
Thus we expect no linear
correction to the simple estimate of $n= N/V$.

\vspace{24pt}
\thesection{\centerline{\bf C. First Order Solution for $w_2(\theta)$}}
\setcounter{section}{3}
\vspace{18pt}

Now, we find the maximum likelihood solution for $w_2$; we will defer
most of the algebraic details to Appendix~A.

First, we will need the derivatives of the correlation function integrals:
\be
{\partial W_0 \over \partial w_2(\theta)} 
= {n^2 V^2\over 2K^2} 2\sum_{i<j} \Theta_{i,j}^\theta
\ee
where $\Theta_{i,j}^\theta$ is one if the distance [in either angular space or
real space depending on whether or not the survey has redshifts]
between cells $i$ and $j$ lies in the bin $\theta$. Using the
definition of Landy \& Szalay, this becomes
\be
{\partial W_0 \over \partial w_2(\theta)} 
\equiv {n^2 V^2\over 2} G_p(\theta) = {n^2 V^2 RR \over N_R^2}
\ee
where $RR$ is the count of pairs at separation $\theta$ in the random
catalogue, as in LS, and $N_R$ is the number of random galaxies
put in; this just normalizes things.
Similarly,
\be
{\partial W_1(x_a) \over \partial w_2(\theta)} 
= - {n V\over K} \sum_{i} \Theta_{i,a}^\theta
\qquad
{\partial W_2(\vx{a},\vx{b}) \over \partial w_2(\theta)} 
= \Theta_{a,b}^\theta
\ee
Another way of viewing the sums over $i,j$ is to think of them
as sums over galaxies in a random catalogue with the same
geometry (and selection function) as the actual survey but with $K$ galaxies
instead of $N$.

We also define the data-data and data-random pair counts as 
\begin{equation}
  DD={1\over2}\sum_a\sum_b^a \Theta_{a,b};\qquad DR=\sum_{aj}\Theta_{a,j}
\end{equation}
where again sums over $a,b,\ldots$ are over observed galaxies and
$i,j,\ldots$ are over cells or the random catalogue.

After quite a bit of algebra (see Appendix A for details), we find
that 
\be\label{lnthree}
{\partial\ln P\over \partial w_2(\theta)} 
= 
{n^2 V^2\over N_R^2}  RR 
 - {n V\over N_R} DR + DD - (DD) w_2(\theta) + E
\ee
where 
\begin{eqnarray}\label{defe}
E &\equiv &
- {n V\over K} \sum_{j,a} w_2(\vx{a},\vx{j})
\Bigg[ {n V\over K} \sum_{i}\Theta_{i,a}^\theta   
-  \sum_b^a 
\Theta_{a,b}^\theta 
\Bigg]\cr
&+&
\sum_a \sum_b^a w_2(\vx{a},\vx{b}) \left[ 
 {n V\over K} \sum_{i} \Theta_{i,a}^\theta 
- \sum_c^b\Theta_{a,c}^\theta \right]
\end{eqnarray}
We reiterate that the sums over $a,b$ indices are over
galaxies in the catalogue, while those over $i,j$ are 
over cells or equivalently over galaxies in a random 
catalogue with the same geometry. In the latter case the number of
cells $K$ can be set to the number of galaxies in the
random catalogue $N_R$.

Equation \ref{lnthree} would lead to a complicated (albeit linear)
matrix equation for 
$w_2$. To simplify, we note that $E$ is usually
small, and is, in fact, negligible to this order in $w_2$, by an
argument similar to that after Eq.~\ref{lnntwo}. To see that this is so,
note that $\sum_b \Theta_{a,b}^\theta$ is simply the number of galaxies
within the bin $\theta$ surrounding the galaxy at $\vx{a}$.
Equivalently, it is $N$ times the fraction of galaxies in the bin
$\theta$ around $\vx{a}$.  If the galaxies were distributed randomly,
this fraction would simply be $(1/K) \sum_i\Theta_{i,a}^\theta$. So the
difference between the two terms in square brackets is due solely to the
non-randomness of the survey, and so is proportional to $w$.  Thus, each
of the terms in square brackets in Eq. \ref{defe}\ are of order
$w_2$; since they multiply terms of
order $w_2$, 
these terms are quadratic in $w_2$. Therefore, they do not contribute at
the level we are interested in.

Finally we are left with the relatively simple expression:
\be\label{lnfinal}
{\partial\ln P\over \partial w_2(\theta)} 
= 
{n^2 V^2\over N_R^2}  RR(\theta) 
 - {n V\over N_R} DR(\theta) + DD(\theta) - DD(\theta) w_2(\theta).
\ee
The likelihood function therefore peaks when
\be
w_2(\theta) = { (n^2 V^2/N_R^2)  RR(\theta) 
 - (n V/N_R) DR(\theta) + DD(\theta) \over DD(\theta)}.
\ee
This expression, while suggestive, is not yet complete.
This expression for $w_2$ depends on the as yet
unknown parameter $n$\footnote{Had we kept the terms in $E$
in our expression for $w$, then the estimate for $w(\theta)$
would depend on $w$ in all other angular bins as well, i.e.
on the other free parameters in the theory. After dispensing
with $E$, we find that the estimate for $w(\theta)$ still
depends on one of the other free parameters in the theory, the
density $n$.}.
As we have seen in the previous subsection, the likelihood function
peaks when $n=N/V$, irrespective of $w(\theta)$; hence we can {\em
  simultaneously} maximize the likelihood for both the density and
correlation function. Inserting this value of $n$ here, we find that
\be
w_2(\theta) = { (N^2/N_R^2)  RR(\theta) 
 - (N/N_R) DR(\theta) + DD(\theta) \over DD(\theta)}.
\ee
This then is the maximum likelihood estimator for $w_2$.
It differs only slightly from the estimator proposed
by Landy \& Szalay; their estimator had $RR$ in the denominator
instead of $DD$. As a point of notation, we mention that LS refer to
their estimator as $(DD-2DR+RR)/RR$; the factor of 2 results from a
normalization choice they have made (in the definition of $d$ in their
Eq.~46). 

In \S IV, we will explore these estimators in
greater detail.

\vspace{24pt}
\thesection{\centerline{\bf D. Width of Likelihood Function}}
\setcounter{section}{3}
\vspace{18pt}

We now want to calculate the width of the likelihood function.
Specifically, we are interested in the matrix
\be
C^{-1}_{\alpha,\beta} \equiv  - {\partial^2 \ln P \over
\partial \lambda_\alpha \partial \lambda_\beta }
\ee
where the parameters in the theory, $\lambda_\alpha$
are the binned $w_2(\theta)$
and $n$. The variances for each of the individual quantities
are then $1/C^{-1}_{\alpha,\alpha}$ if all other parameters
are held fixed, while $C_{\alpha,\beta}$ is the general covariance matrix
if all parameters are allowed to vary.

Let us first calculate $C^{-1}_{n,n}$ by
differentiating Eq.~\ref{lnntwo}. We find
\be
C^{-1}_{n,n} = {N\over n^2} - {V^2\over K^2} \sum_{i,j} w_2(\vx{i},\vx{j})
.\ee
So the variance in $n$ is increased by the presence of correlations.

The next element is obtained by differentiating Eq.~\ref{lnntwo}
with respect to $w_2$. We find
\be
C^{-1}_{n,w_2(\theta)} = 
2 nV^2 {RR\over N_R^2} - 
 {V\over N_R}DR
\ee
Note that this vanishes if there are no correlations. Note also
that since Eq.~\ref{lnntwo} was accurate only up to order $w_2$,
this expression is accurate only up to order $w_2^0$.
This is true for $C^{-1}_{w_2,w_2}$ as well: since 
our expression for $\partial \ln P/\partial w_2$ is accurate
only up to order $w_2$, the second derivative of
will not contain any information about the $w_2$
dependence of the width. We will not be able to distinguish between
the width due to a random catalogue from that due
to a correlated one.
Carrying through this differentiation on Eq.~\ref{lnfinal}, we find
\be\label{lastel}
C^{-1}_{w_2(\theta_1),w_2(\theta_2)} =  DD(\theta_1) \delta_{\theta1,\theta2}
\ee
where $\delta_{\theta1,\theta2}$ is the Kronecker delta.
This agrees with Landy \& Szalay's calculation, neglecting
corrections of order $w_2$. Correlations between $w_2$'s at different
$\theta$'s would appear in the higher order terms. 
To assess the relative effectiveness
of these two estimators, we could go back to our expansion
and attempt to extract the order $w_2^3$ terms. Alternatively, we could
perform a Monte Carlo to see which has the lower variance. 
We choose the latter option.

\vspace{48pt}
\thesection{\centerline{\large \bf IV. Estimators of the Two Point
Function}}
\setcounter{section}{4}
\vspace{18pt}

Here we would like to test the effectiveness of various
estimators. Before presenting our results, we briefly review
previous work. Landy \& Szalay analyzed random catalogues,
attempting to measure the variance of their estimator and the more
traditional $DD/DR$. They found that the variance of their
estimator was very close to the expected Poisson value:
\be\label{POISSON}
\sigma^2_{\rm Poisson} = {N_R^2\over N^2 RR(\theta)}
.\ee
Note that this is simply one over the expected number of galaxy pairs
per bin squared, $1/N_{\rm pairs}^2$
On the other hand, the
$DD/DR$ estimator gave a larger variance than this at 
large angles. They attributed this to the fact that at large
distances, the number of galaxy pairs per bin goes up and
the $DD/DR$ variance has an additional term beyond Poisson
which goes as $1/N_{\rm pairs}$. As $N_{\rm pairs}$ gets larger,
this additional term eventually dominates. Bernstein (1993)
simulated
catalogues with non-zero $n$-point correlations. He found that
the Landy \& Szalay estimator had a larger variance in this
correlated case than what one would expect based on Poisson-counting.
As we will see, our work agrees with 
both of these results.

We work in a $32^2$ box (the units are irrelevant).
An outline of our recipe is:

\begin{enumerate}

\item Generate a galaxy catalogue with of order $1500$ galaxies.

\item Compute the expected value of $w_2$ from this catalogue using
three different estimators: Landy-Szalay (LS), $DD/DR$, and
our maximum likelihood (ML).

\item Repeat steps 1 \& 2, $200$ times.

\item Calculate the mean and variance of these estimators over the
  ensemble of realizations.

\end{enumerate}

To generate a galaxy catalogue (step 1), we input our desired
$w_2(\theta)$, and Fourier transformed to get the power spectrum.
We then used this power spectrum to generate a density field 
everywhere on a $32^2$ grid. We chose the amplitude of $w_2$ to
be small enough so that $\delta$ was never less than $-1$ anywhere
on the grid. Then, we used the density field to produce a Poisson
realization with at most ten galaxies per cell. Then we randomized the
positions within each cell. Several comments about this proceedure:
First, we are by necessity limited to small $w_2$. Second, we do
not trust our results on scales smaller than one cell size ($x=1$).
Third, because of periodicity, $w_2$ obtained in this way
is symmetric about the half-way point ($x=16$), so we restrict our
analysis to $x<16$.

To calculate the expected value of $w_2$ (step 2), we generate thirty
random catalogues with one thousand particles each.\footnote{This should
be equivalent to one catalogue of $30,000$ particles but is computationally
faster to analyze. We have checked that this is sufficient by
calculating the expected variance of $w_2$ for a random catalogue \`a la
Landy \& Szalay. We agree with the expected variance in that case.}
We calculate $DD,DR,RR$ in each of seven bins, the
lowest at $1<x<3$ and the highest at $13<x<15$. From these, we
construct the three estimators of interests.

\begin{figure}
\centerline{ \epsfxsize=400pt \epsfbox{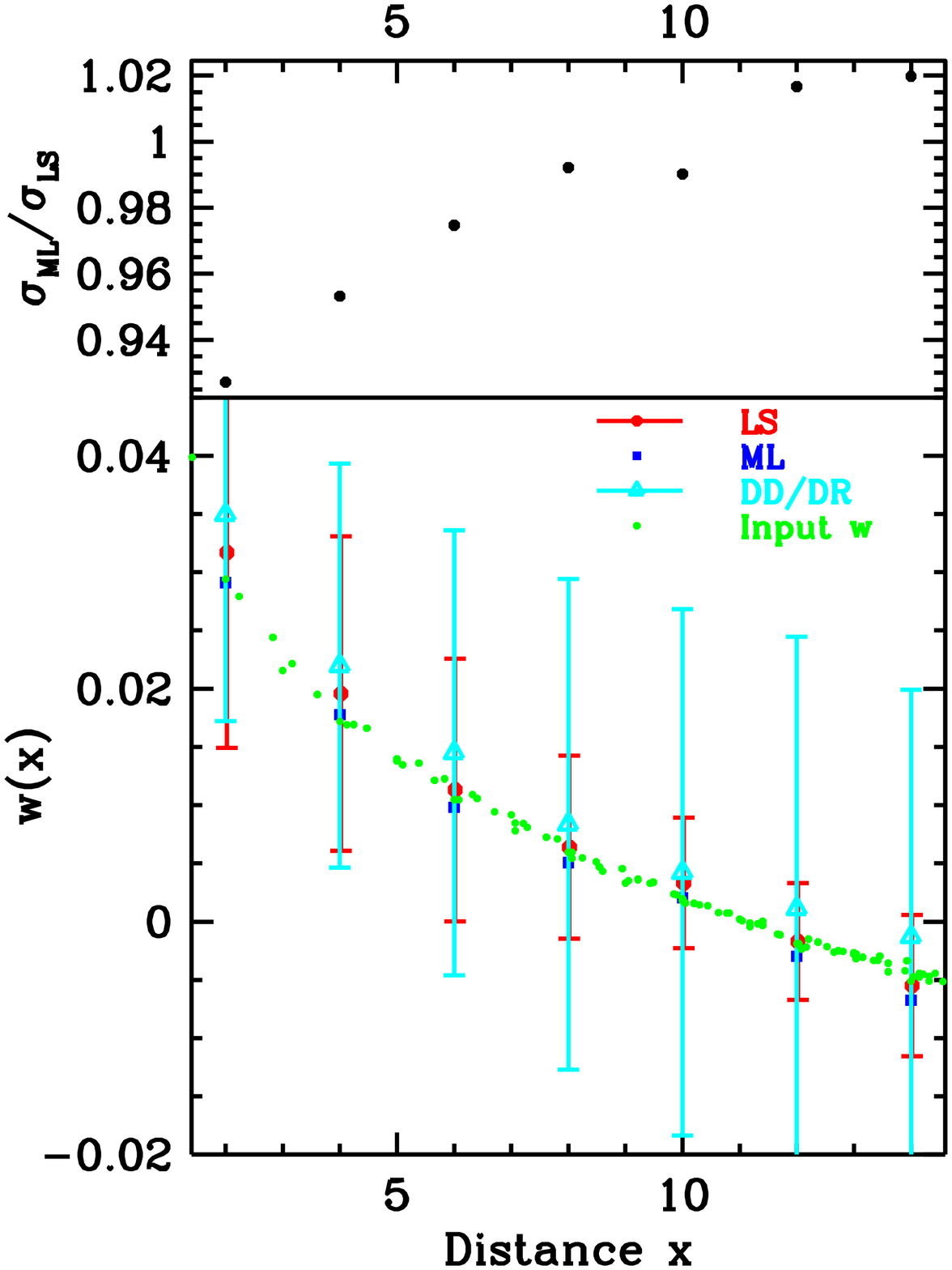} }
\footnotesize{\hspace*{0.2in} Fig.\ 1:  The bottom panel shows the input
$w_2$ and the estimated values, using three different estimators.
The error bars for the Landy \& Szalay estimator are plotted
along with those for $DD/DR$. At large distance (with many galaxy
pairs per bin) the Landy \& Szalay estimator has significantly
smaller variance than does $DD/DR$. The error bars for our maximum likelihood
estimator are similar to those of the Landy \& Szalay estimator
so are not shown explicitly in the bottom panel. 
The ratio of the two is shown in the top panel. For bins
with non-negligible
$w_2$, the maximum likelihood estimator appears to have a smaller
variance.}
\end{figure}

Our results are shown in Figures 1 and 2. Figure 1 shows that all of
the estimators come very close to the true value of $w_2$. 
The variance of the LS estimator is indeed significantly smaller
than $DD/DR$ at large separations when the $1/N_{\rm pairs}$
factor begins to overtake the Poisson variance. The variance of
the ML estimator is similar to LS, but as shown in the top panel
of Figure 1, appears to be smaller when $w_2$ is non-negligible.

\begin{figure}
\centerline{ \epsfxsize=400pt \epsfbox{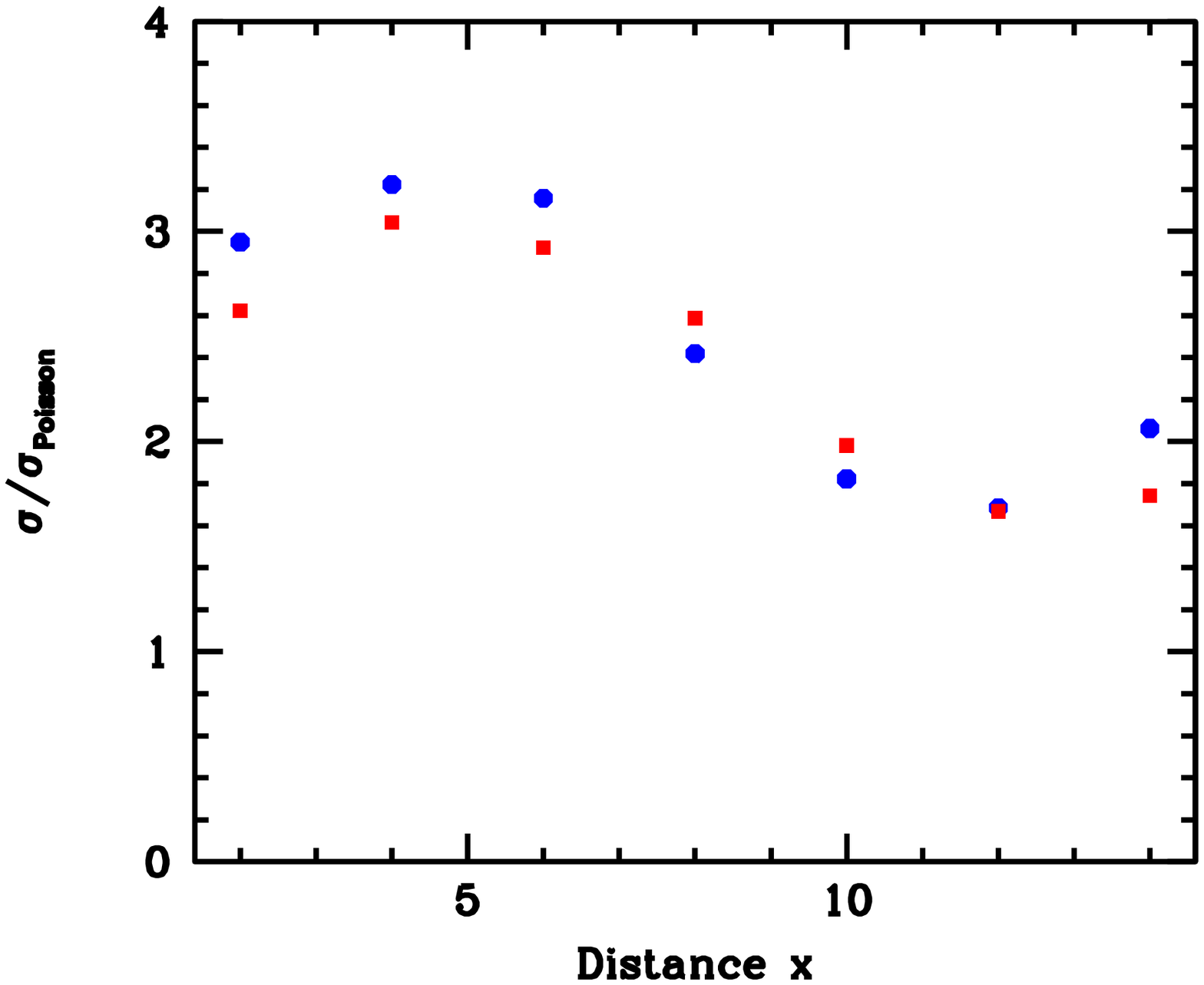} }
\footnotesize{\hspace*{0.2in} Fig.\ 2:  The ratio of the
rms variation of the Landy \& Szalay estimator
to the ``expected'' Poisson rms, $\sqrt{N_R^2/(N^2 RR)}$.
When $w_2$ is non-negligible (at small $x$) the rms
variation is significantly larger than Poisson. (Blue)
circles are variance obtained from 200 simulated
catalogues;
(red) squares are the variance expected from analytic
results in Appendix B.}
\end{figure}
There is one other feature of our analysis which bears note. 
Figure 2 shows the variance of the LS estimator as compared with
the ``expected'' Poisson variance, Eq.~\ref{POISSON}. The variance of
the LS estimator
is significantly larger when $w_2$ is non-negligible. 
We believe this is real and that there are two non-negligible additional
terms in the LS 
variance, one proportional to $w_2/N$, and one to $w_2^2$. Appendix B
derives these additional 
terms analytically. Both the numerical and the analytic
result agree with Bernstein's results. The differences are: we have
included edge effects and we have isolated the fact that the
discrepency is due to these added terms in the variance. It is not
due to higher order cosmic correlations, as these are excluded by
construction in our mock catalogue.

\vspace{48pt}
\thesection{\centerline{\large \bf V. Conclusions and the Future}}
\setcounter{section}{5}
\vspace{18pt}

The likelihood function would be a wonderful object to compute for a
galaxy survey. By construction, it would use {\rm all} of the
information from the survey and allow us to compare different theories.
As such, we feel it is very important to explore the possibility of
computing this function. Direct computation is impossible, but we have
developed an approximation scheme which appears to work very well.
Before specifically detailing the use we have made of this
approximation, we want to emphasize that there are many ways to branch
out from here:

\begin{itemize}

\item Use a theory such as Cold Dark Matter with its one or two free
parameters. The approximation scheme developed here can then be used
to extract the best fit values of these parameters.

\item Generalize the approximation to include higher-order correlations.
This would have the benefit of using the higher-order correlations together
with the two-point function to constrain theories. Alternately, we could
use the ansatz of hierarchical clustering and results from gravitational
perturbation theory to generate higher-order moments from the two-point
function.

\item Generalize this work to Fourier space. Theories are most
easily compared in Fourier space so this is a natural way to go. Just as
the ML procedure generates a (diagonal) matrix equation for $w_2$, in
Fourier space we have an integral equation for $P(k)$.

\item Find a graphical method which simplifies, and helps organize,
the expansion we have introduced. We have relied on arguments
like: certain terms, such as $E$ (Eq. \ref{defe}), are implicitly of
the order 
of ${w_2}^2$, even though they 
do not appear explicitly so. It would be nice to 
make these more precise and systematic. 

\item Go beyond the perturbative approach and try to learn something
from the full likelihood function. This is not as impossible as we
made it sound in \S II: without actually computing $\like$, one
might still make some very general statements.

\end{itemize}

We have made progress in the latter four areas. 
This will be presented in a future paper.

In this paper we have limited ourselves to one application: finding
the place where the likelihood function peaks if the theoretical
parameters are the binned values of the two-point function. Equivalently,
we have come up with a new estimator for $w_2$. We found that

\begin{itemize}

\item The maximum likelihood (ML) estimator appears to have a slightly smaller variance than
the Landy \& Szalay (LS) estimator and certainly than $DD/DR$. To the extent that 
the the ML and LS estimators are similar (and they are {\em very} similar) this whole
treatment can be thought of as further motivation for the LS estimator. 

\item There are additional terms in the variance of the LS estimator beyond the
Poisson variance. These terms begin to dominate when correlations are
non-negligible 
and the number of pairs of galaxies per bin is large. 

\end{itemize}

\vspace{36pt}

This work was supported in part by DOE and NASA grant NAG5--2788 at Fermilab.
We thank Istvan Szapudi for useful conversations.


\begin{picture}(400,50)(0,0)
\put (50,0){\line(350,0){300}}
\end{picture}

\vspace{0.25in}

\def\labelenumi{[\theenumi]}
\frenchspacing
\def\prl{{{\em Phys. Rev. Lett.\ }}}
\def\prd{{{\em Phys. Rev. D\ }}}
\def\pl{{{\em Phys. Lett.\ }}}

\begin{enumerate}

\item\label{b93} G. M. Bernstein,
Astrophysical Journal {\bf 424}, 569 (1994).

\item\label{hamilton93} A. J. S. Hamilton,
Astrophysical Journal {\bf 417}, 19 (1993).

\item\label{hamilton97} A. J. S. Hamilton,
Monthly Notices of Royal Astronomical Society {\bf 289}, 285 (1997).

\item\label{hewett} H.C. Hewett,
Monthly Notices of Royal Astronomical Society {\bf 201}, 867 (1982).

\item\label{al93} S. D. Landy \& A. S. Szalay,
Astrophysical Journal {\bf 412}, 64 (1993).

\item\label{MJB} H. J. Mo, Y.P. Jing, \& G. B\"{o}rner,
Astrophysical Journal {\bf 392}, 452 (1992).

\item\label{peebles} P.J.E. Peebles, The Large Scale Structure
of the Universe (Princeton: Princeton University Press) (1980).

\item\label{ss} I. Szapudi \& A. S. Szalay, astro-ph/9704241 (1997).

\item\label{white79} S. D. M. White,
Monthly Notices of Royal Astronomical Society {\bf 186}, 145 (1979).

\end{enumerate}

\newpage

\vspace{48pt}
\thesection{\centerline{\large \bf Appendix A. Derivation
of Weak Correlation Limit}}
\setcounter{section}{1}
\renewcommand{\ksection}{\Alph{section}}
\renewcommand{\theequation}{A\arabic{equation}}
\setcounter{equation}{0}
\vspace{18pt}
We will take derivatives of Eq.~\ref{lnone}\ with respect to $w_2$; and
then expand it out to first order in $w_2$. 
\begin{eqnarray}{\partial\ln P\over \partial w_2(\theta)} &&
= {n^2 V^2 RR \over N_R^2} 
 \cr
+
&& \Bigg[ \prod_a W_1(\vx{a})   
+  {1\over 2} \sum_a \sum_b^a 
W_2(\vx{a},\vx{b}) \prod_c^{a,b} W_1(\vx{c})  \Bigg]^{-1} \cr
\times &&
\Bigg[ - {n V\over K} \sum_{b,i}\Theta_{i,b}^\theta\prod_a^b W_1(\vx{a})   
+ {1\over 2} \sum_a \sum_b^a 
 {\partial\over \partial w_2(\theta)}\Big\{
W_2(\vx{a},\vx{b}) \prod_c^{a,b} W_1(\vx{c})\Big\}  \cr
+ && {1\over 2^3} \sum_a \sum_b^a \sum_c^{a,b} \sum_d^{a,b,c}
{\partial\over \partial w_2(\theta)}\Big\{
W_2(\vx{a},\vx{b}) W_2(\vx{c},\vx{d}) \prod_e^{a,b,c,d} W_1(\vx{e})
\Big\}\Bigg]
\end{eqnarray}
Since we are keeping only first order terms here, we have dropped the 
$W_2^2$ term in the denominator. We can go further though. The term in the
denominator linear in $W_2$ is multiplied by the product of $W_1$'s; these
can all be set to one. Similarly, the derivative operator acting on
the last term only affects the $W_2$'s; the $W_1$'s can again all be set
to one. 
This gives:
\begin{eqnarray}{\partial\ln P\over \partial w_2(\theta)} 
&=& {n^2 V^2 RR \over N_R^2} 
 \cr
&+
& \Bigg[ \prod_a W_1(\vx{a})   
+  {1\over 2} \sum_a \sum_b^a 
W_2(\vx{a},\vx{b})  \Bigg]^{-1} \cr
&\times &
\Bigg[ - {n V\over K} \sum_{b,i}\Theta_{i,b}^\theta\prod_a^b W_1(\vx{a})   
+ {1\over 2} \sum_a \sum_b^a 
 {\partial\over \partial w_2(\theta)}\Big\{
W_2(\vx{a},\vx{b}) \prod_c^{a,b} W_1(\vx{c})\Big\}  \cr
&+ & {1\over 2^3} \sum_a \sum_b^a \sum_c^{a,b} \sum_d^{a,b,c}
{\partial\over \partial w_2(\theta)}\Big\{
W_2(\vx{a},\vx{b}) W_2(\vx{c},\vx{d}) 
\Big\}\Bigg]
\end{eqnarray}
Now carry out the derivatives:
\begin{eqnarray}{\partial\ln P\over \partial w_2(\theta)} 
&=& {n^2 V^2 RR \over N_R^2} 
 \cr
&+&
 \Bigg[ \prod_a W_1(\vx{a})   
+  {1\over 2} \sum_a \sum_b^a 
W_2(\vx{a},\vx{b})  \Bigg]^{-1} \cr
&\times &
\Bigg[ - {n V\over K} \sum_{b,i}\Theta_{i,b}^\theta\prod_a^b W_1(\vx{a})   
+ {1\over 2} \sum_a \sum_b^a 
 \Big( \Theta_{a,b}^\theta
\prod_c^{a,b} W_1(\vx{c})
\cr
&-& {n V\over K}  w_2(\vx{a},\vx{b}) \sum_e^{a,b} \sum_{i} \Theta_{i,e}^\theta
 \Big)
  \cr
&+ & {1\over 2^3} \sum_a \sum_b^a \sum_c^{a,b} \sum_d^{a,b,c}
\left(\Theta_{a,b}^\theta w_2(\vx{c},\vx{d}) 
+ \Theta_{c,d}^\theta w_2(\vx{a},\vx{b})  \right)
\Bigg]
\end{eqnarray}
Now expand the denominator:
\be
{1\over \prod_a W_1(\vx{a})   
+  {1\over 2} \sum_a \sum_b^a 
W_2(\vx{a},\vx{b}) } = 
1 - {1\over 2} \sum_a \sum_b^a 
w_2(\vx{a},\vx{b}) 
- \sum_a [W_1(\vx{a})- 1]
\ee

So,
\begin{eqnarray}{\partial\ln P\over \partial w_2(\theta)} &&
= {n^2 V^2\over N_R^2}  RR 
 \cr
+
&& \left[ 1 - {1\over 2} \sum_a \sum_b^a 
w_2(\vx{a},\vx{b}) 
- \sum_a (W_1(\vx{a})- 1) \right]\cr
\times &&
\Bigg[ - {n V\over K} \sum_{b,i}\Theta_{i,b}^\theta\prod_a^b W_1(\vx{a})   
+ {1\over 2} \sum_a \sum_b^a 
 \Big( \Theta_{a,b}^\theta
\prod_c^{a,b} W_1(\vx{c})
\cr
&&
- {n V\over K}  w_2(\vx{a},\vx{b}) \sum_e^{a,b} \sum_{i} \Theta_{i,e}^\theta
 \Big)
  \cr
+ && {1\over 2^3} \sum_a \sum_b^a \sum_c^{a,b} \sum_d^{a,b,c}
\left(\Theta_{a,b}^\theta w_2(\vx{c},\vx{d}) 
+ \Theta_{c,d}^\theta w_2(\vx{a},\vx{b})  \right)
\Bigg] 
\end{eqnarray}
Multiplying through, we find
\begin{eqnarray}\label{lntwo}
{\partial\ln P\over \partial w_2(\theta)} &&
= {n^2 V^2\over N_R^2}  RR 
 - {n V\over K} \sum_{b,i}\Theta_{i,b}^\theta\prod_a^b W_1(\vx{a})   
+ {1\over 2} \sum_a \sum_b^a 
 \Big( \Theta_{a,b}^\theta
\prod_c^{a,b} W_1(\vx{c})
\cr
&&
- {n V\over K}  w_2(\vx{a},\vx{b}) \sum_e^{a,b} \sum_{i} \Theta_{i,e}^\theta
 \Big)
  \cr
+ && {1\over 2^3} \sum_a \sum_b^a \sum_c^{a,b} \sum_d^{a,b,c}
\left(\Theta_{a,b}^\theta w_2(\vx{c},\vx{d}) 
+ \Theta_{c,d}^\theta w_2(\vx{a},\vx{b})  \right)
\cr
&&- \left[ {1\over 2} \sum_a \sum_b^a 
w_2(\vx{a},\vx{b}) 
+ \sum_a (W_1(\vx{a})- 1) \right]
\cr &&
\times
\Bigg[ - {n V\over K} \sum_{b,i}\Theta_{i,b}^\theta   
+ {1\over 2} \sum_a \sum_b^a 
\Theta_{a,b}^\theta
\Bigg]
\end{eqnarray}
There are three sets of terms here, those which have $w_2$
explicitly in them; those which are independent of $w_2$;
and those which depend on $w_2$ only through $W_1-1$. 
Let us treat each of these in turn.

First consider the terms independent of $w_2$ in Eq.~\ref{lntwo}:
\be
{n^2 V^2\over N_R^2}  RR 
 - {n V\over K} \sum_{b,i}\Theta_{i,b}^\theta  
+ {1\over 2} \sum_a \sum_b^a 
 \Theta_{a,b}^\theta
= {n^2 V^2\over N_R^2}  RR 
 - { n V\over N_R} DR
+ DD
\ee

Next consider the terms which depend explicitly on $w_2$.
\begin{eqnarray}
- {n V\over 2K} \sum_a \sum_b^a w_2(\vx{a},\vx{b}) \sum_e^{a,b} \sum_{i} \Theta_{i,e}^\theta
&+&  {1\over 2^3} \sum_a \sum_b^a \sum_c^{a,b} \sum_d^{a,b,c}
\left(\Theta_{a,b}^\theta w_2(\vx{c},\vx{d}) 
+ \Theta_{c,d}^\theta w_2(\vx{a},\vx{b})  \right)
\cr
- {1\over 2} \sum_a \sum_b^a 
w_2(\vx{a},\vx{b}) 
\Bigg[ - {n V\over N_R} DR 
&+& DD
\Bigg]
\end{eqnarray}
We claim that the two terms in the quadruple sum on the
first line are identical. To see this, first switch indices
$a\leftrightarrow c; b\leftrightarrow d$ in the first term. Then it is:
\be 
\sum_c \sum_d^c \sum_a^{c,d} \sum_b^{c,d,a}
\Theta_{c,d}^\theta w_2(\vx{a},\vx{b})
= \sum_a^{c} \sum_c^a  \sum_d^{a,c}  \sum_b^{c,d,a}
\Theta_{c,d}^\theta w_2(\vx{a},\vx{b})
\ee
where we have switched the summations and taken care to
guard against summing over identical $a=c$ for example.
Continuing in this fashion, we get the second term in
the quadruple sum. Thus, all terms with explicit
$w_2$'s in them are:
\be
\sum_a \sum_b^a w_2(\vx{a},\vx{b}) \left[ 
- {n V\over 2K} \sum_e^{a,b} \sum_{i} \Theta_{i,e}^\theta
+ {1\over 2^2} \sum_c^{a,b} \sum_d^{a,b,c}
\Theta_{c,d}^\theta 
- {1\over 2} 
\Big( - {n V\over N_R} DR 
+ DD
\Big) \right]
\ee
Consider the first term in brackets. We can rewrite
the sum over $e$ as 
\be
\sum_e^{a,b} = \sum_e \left(1-\delta_{ae}\right)\left(1-\delta_{be}\right)
 = \sum_e \left(1- \delta_{ae} - \delta_{be}\right)
\ee
Note that since $a$ is never equal to
$b$, this is exactly true. Now the unrestricted sum
over $e$ simply gives $DR$ with the sign and coefficient
exactly right to cancel the third term in brackets.
So the terms with explicit $w_2$ dependence are:
\be
\sum_a \sum_b^a w_2(\vx{a},\vx{b}) \left[ 
 {n V\over 2K} \sum_{i} (\Theta_{i,a}^\theta + \Theta_{i,b}^\theta)
+ {1\over 2^2} \sum_c^{a,b} \sum_d^{a,b,c}
\Theta_{c,d}^\theta 
- {1\over 2} 
DD
\right]
\ee
The same argument holds for the $DD$ terms. The only terms which
remain in the sums over $c,d$ are those wherein $c$ or $d$ equals
$a,b,c$. Let us do this carefully because there might be subtleties
here.
\begin{eqnarray}
\sum_c^{a,b} \sum_d^{a,b,c}
&=& \sum_c^{a,b}\sum_d^{c}\left( 1 - \delta_{da} -\delta_{db}\right)
\cr
&=& \sum_c\sum_d^{c}\left[ 1  - 
\left(\delta_{ac} + \delta_{bc} \right)\right]
- \sum_c^{a,b}\sum_d (\delta_{da} +\delta_{db} )
\end{eqnarray}
So the explicit $w_2$ terms become
\be
\sum_a \sum_b^a w_2(\vx{a},\vx{b}) \left[ 
 {n V\over 2K} \sum_{i} (\Theta_{i,a}^\theta + \Theta_{i,b}^\theta)
- {1\over 2^2} [\sum_d^a\Theta_{a,d}^\theta 
+\sum_d^b\Theta_{b,d}^\theta
+ \sum_c^{a,b} ( \Theta_{c,a}^\theta + \Theta_{b,c}^\theta)]
\right]
\ee
We want to focus on one of these terms: the one in which the index $d$
is equal to $a$ or $b$. Thus rewrite this last line as
\be
\sum_a \sum_b^a w_2(\vx{a},\vx{b}) \left[ 
 {n V\over K} \sum_{i} \Theta_{i,a}^\theta 
- \sum_d^b\Theta_{a,d}^\theta \right]
- {1\over 2} \sum_a\sum_b^a w_2(\vx{a},\vx{b}) \Theta_{a,b}^\theta
\ee
where we have made use of the symmetry between $a,b$ to sum
up lots of identical terms. Now consider the last term. The
$ \Theta_{a,b}^\theta$ requires all separations between $\vx{a}$
and $\vx{b}$ to lie within the bin $\theta$. Within this bin, by
definition, $w_2$ is constant [this is the theory we are trying
to solve for]. Thus, $w_2$ comes out of the sum and we are left
with simply $DD$. So the terms with only explicit $w_2$ dependence
are
\be
\sum_a \sum_b^a w_2(\vx{a},\vx{b}) \left[ 
 {n V\over K} \sum_{i} \Theta_{i,a}^\theta 
- \sum_d^b\Theta_{a,d}^\theta \right]
- (DD) w_2(\theta)
\ee

We can now reinsert all this back into Eq. \ref{lntwo}.
\begin{eqnarray}
{\partial\ln P\over \partial w_2(\theta)} &
=& 
{n^2 V^2\over N_R^2}  RR 
 - {n V\over N_R} DR + DD - (DD) w_2(\theta)
\cr
 &-& {n V\over K} \sum_{b,i}\Theta_{i,b}^\theta\sum_a^b [W_1(\vx{a})-1]   
+ {1\over 2} \sum_a \sum_b^a 
 \Theta_{a,b}^\theta
\sum_c^{a,b} [W_1(\vx{c})-1]
\cr
&-&
 \sum_c (W_1(\vx{c})- 1)
\Bigg[ - {n V\over K} \sum_{b,i}\Theta_{i,b}^\theta   
+ {1\over 2} \sum_a \sum_b^a 
\Theta_{a,b}^\theta
\Bigg]\cr
&+&
\sum_a \sum_b^a w_2(\vx{a},\vx{b}) \left[ 
 {n V\over K} \sum_{i} \Theta_{i,a}^\theta 
- \sum_d^b\Theta_{a,d}^\theta \right]
\end{eqnarray}
The terms in the middle two lines nearly cancel, except for 
the restrictions on the sums. The only terms that remain
from these two lines are those in which $c$ on the third line equals $a$ or
$b$. Thus we are lead directly to Eq.~\ref{lnthree}.

\vspace{48pt}
\thesection{\centerline{\large \bf Appendix B. Variance of the
Landy \& Szalay Estimator}}
\setcounter{section}{2}
\renewcommand{\ksection}{\Alph{section}}
\renewcommand{\theequation}{B\arabic{equation}}
\setcounter{equation}{0}
\vspace{18pt}

To derive the variance of the Landy \& Szalay estimator, let us rewrite
it in the following form:

\begin{equation}
w_2^{\rm LS} (\theta) = \sum_{i,j} W_{i,j}(\theta)  ({q_i- \alpha
q^r_i}) ({q_j- \alpha q^r_j}) \, ,
\label{LSalternate}
\end{equation}
where 
\begin{equation}
W_{i,j}(\theta) = {\Theta_{i,j}(\theta) \over \sum_{m,n}
\Theta_{m,n}(\theta) \alpha^2 q^r_m q^r_n} \, ,
\label{WLS}
\end{equation}
and $q_i$ equals one where the cell contains a galaxy, and equals zero
otherwise, whereas $q^r_i$ is similarly defined for the random
catalogue. 
The factor $\alpha$ scales the average density in random catalogue back
to the level in the actual catalogue: in other words, $\alpha = N/N_R$
with $N$ and $N_R$ being the total number of galaxies in the actual
catalogue and in the random catalogue respectively. 

It can be shown that the introduction of the random catalogue
introduces extra variance (i.e. variance of the random catalogue
itself), which could be eliminated if one uses a large number of random
catalogues, or if one allows the number of galaxies in random catalogue
to increase dramatically. 
We will compute the variance of $w_2^{\rm LS} (\theta)$ in this limit,
in which case, one can replace every $\alpha q^r_i$ by $\bar q$ where
$\bar q = \langle q_i \rangle$. Hence, Eq.~\ref{LSalternate} is
equivalent to 
\begin{equation}
w_2^{\rm LS} (\theta) = \sum_{i,j} \tilde W_{i,j}(\theta) \delta_i
\delta_j
\label{wLSalternate}
\end{equation}
with
\begin{equation}
\tilde W_{i,j}(\theta) = {\Theta_{i,j}(\theta) \over \sum_{m,n}
\Theta_{m,n}(\theta)}
\label{tWLS}
\end{equation}
and $\delta_i = (q_i - \bar q)/\bar q$. 

The variance is defined by $\langle [w_2^{\rm LS} (\theta)]^2 \rangle
- \langle w_2^{\rm LS} (\theta) \rangle^2$. This means one needs
the following quantity:
\begin{eqnarray}
\label{Delta}
&& \langle \delta_i \delta_j \delta_k \delta_l \rangle - \langle \delta_i
\delta_j \rangle \langle \delta_k
\delta_l \rangle  \\ \nonumber 
&& = [w_2(\vx{i},\vx{k}) + \delta_{ik} \bar q_i
{^{-1}}] [w_2(\vx{j},\vx{l}) + \delta_{jl} \bar q_i
{^{-1}}] + (k \leftrightarrow l)
\end{eqnarray}
which is adopted from Hamilton (1997). This expression holds
in the small pixel (continuum) limit, with the restriction $i \ne j$
and $k \ne l$ (for nonzero $\theta$) and for Gaussian random
underlying field. We have also ignored terms the contributions of
which to the variance are of order of $1/N$
smaller than those we have kept. 

Putting everything together, we obtain:
\begin{eqnarray}
\label{LSvariance}
{\rm Variance} &&= {2 \over {\bar q^2 \sum_{i,j}\Theta_{i,j}(\theta)}} +
{4 \sum_{i,j,l} \Theta_{i,j}(\theta) \Theta_{i,l}(\theta)
w_2(\vx{j},\vx{l}) \over {\bar q [\sum_{m,n} \Theta_{m,n}(\theta)]^2}}
\\ \nonumber
&& + {2 \sum_{i,j,k,l} \Theta_{i,j}(\theta) \Theta_{k,l}(\theta)
w_2(\vx{i},\vx{k}) w_2(\vx{j},\vx{l}) \over {[\sum_{m,n}
\Theta_{m,n}(\theta)]^2}} 
\end{eqnarray}

The first term on the right is the Poisson variance given by Landy \&
Szalay (1993). The second two terms arise because of finite
two-point correlations. They have been derived before by Bernstein
(1993), ignoring edge effects, and by Mo, Jing, \& B\"{o}rner,
ignoring discreteness effects. Note that both Landy \& Szalay and
Bernstein obtained extra terms for the Poisson variance, which we have
ignored because they are of the order of $1/N$ smaller than those we
have kept. The variance in Eq. \ref{LSvariance} can be
estimated using the standard 
technique of counting pairs, triplets and quadruplets using a random
catalogue with the same geometry. For instance one can estimate
$\bar q^2 \sum_{i,j}\Theta_{i,j}(\theta)/2$ by $[N^2 / N_R^2] RR(\theta)$.

\end{document}